# Facets of Brachistochronic Trajectories

"The task is not only to see what no one has seen before but, more importantly, to think thoughts that no one has thought before about things that everyone has already seen." (Erwin Schrödinger)

David Agmon, agmon1@campus.technion.ac.il and Ady Mann, ady@physics.technion.ac.il

## Abstract

Given a conservative force field, a necessary and sufficient condition for a curve to be a brachistochrone is that the magnitude of the normal component of the conservative force equals that of the centrifugal force: $F_n = mv^2/R_c$, where $R_c$ denotes the radius of curvature. This leads to two general rules valid for any brachistochrone under an arbitrary conservative force:

**1a. The rule of 2**: The normal force exerted by the guiding wire is always twice the magnitude of the normal component of the conservative force and opposite in direction.

**1b. Mirror rule**: The net force and the conservative force have equal magnitudes, with the net force being the mirror image of the conservative force with respect to the tangent to the path.

In the specific case of a central force, the following two equivalent rules also hold:

**2a.** The ratio of angular momentum to kinetic energy remains constant along the trajectory.

**2b.** The time taken to travel from the center of force to the tangent line is constant (see Fig. 2).

We establish these results using two complementary approaches. The first exploits the fact that any brachistochrone trajectory must be stationary. The second directly applies the Euler–Lagrange (E–L) equation.

Finally, we show that for a *central* force, the guiding wire can be replaced by a suitable magnetic field. This substitution eliminates the need for the E–L formalism, allowing the problem to be solved by directly integrating the equations of motion for a charged particle in combined electric and magnetic fields.

## Preamble

Our interest in the *general* brachistochrone problem arose from a discrete approach to the classical Bernoulli brachistochrone [1, 2], which revealed several intriguing properties. Upon further investigation of the discrete formulation (to be presented in a forthcoming publication), we discovered that some of these properties are, in fact, quite general and arise from the fundamental condition of stationarity. This observation motivated us to study the general case of stationarity in the continuous brachistochrone problem. Our analysis uncovered several noteworthy features that, to the best of our knowledge, are absent from the literature of the 20th and 21st centuries. Interestingly, we later found that some of these results were already mentioned in Routh's 1898 treatise [3]—with some apparently known to Euler as early as 1736 [3]. Since our derivations are different in method and since this topic appears to be largely overlooked in the modern literature, we believe it merits renewed attention. Notably, our results extend naturally to the framework of special relativity, which, of course, could not have been considered prior to 1905.

## Introduction

When searching for a brachistochronic trajectory, the simplest case is to determine the equation of a path connecting two given points in a region where a prescribed conservative force acts, such that the travel time of a bead sliding along it is minimized. Consequently, such a trajectory represents an optimal compromise between two competing requirements: minimizing the path length while maximizing the average velocity.

The starting point of our research was actually based on a discrete approach [1,2] in which it is assumed that the brachistochronic trajectory consists of tiny, straight and smooth segments and the transition between them is smooth and energy conserving. The results were quite surprising. E.g. the sliding times along all segments are equal. These results were first obtained for the "classical brachistochrone" of Bernoulli and we found later that they hold for a general *central* force and this encouraged us to employ the stationary approach. We start with the stationary approach based on the fact that *any* infinitesimal first-order variation in the length of the stationary path induces only a second-order or higher variation in the total gliding time along the path. Using this fact allows us to easily determine the general properties of brachistochronic trajectories without relying on the Euler-Lagrange (E-L) equation.

Two forces control this motion: the conservative force $\vec{F}$ and a steering force $\vec{N}$.

$\vec{N}$ is the normal force exerted by the smooth wire on the bead threaded on it. Since *all* brachistochronic trajectories are stationary, it is expected that they share some common features, regardless of the particular form of the force that governs their motion. For example, it turns out that for a brachistochrone under a *central* force the quantity $L_0 / E_k$ is conserved, where $L_0$ is the angular momentum and $E_k$ is the kinetic energy.

A necessary and sufficient condition for brachistochrone motion is the requirement that the normal component of the conservative force equals the centrifugal force on the bead. From this condition, we derive two equivalent rules, each of which may independently serve as a necessary and sufficient condition for the existence of brachistochrone motion:

1a. "The rule of 2" states that the steering force $\vec{N}$ is twice the size of the normal component of the conservative force and opposite in direction: $\vec{N} = -2\vec{F}_n$.

1b. "The mirror rule" states that the *net* force on the bead, $\vec{f}_{net} \equiv \vec{F} + \vec{N}$, and the original conservative force $\vec{F}$ have the same magnitude ($|\vec{f}_{net}| = |\vec{F}|$) *and* the net force is a mirror image of the conservative force reflected through the tangent to the brachistochronic trajectory. These results are shown in Fig. 1.

When the conservative force is a *central* force, two additional interwoven rules emerge:

2a. The ratio of angular momentum to kinetic energy is constant: $(L_o / E_k) = 2T_0$.

2b. The time required to travel from the center of force to the tangent line is constant. I.e. the expression $r \sin \alpha / v \equiv T_0$ (cf. Fig. 2.) is constant.

Note: Our results were also extended to the relativistic case (to be addressed separately), in which both proper time and laboratory time are relevant. For the proper time, the results are analogous to those of the nonrelativistic case. However, for the laboratory time, we derived an intriguing relation $N / F_n = (2 - \beta^2)$, where $\beta \equiv v / c$.

In the classical limit $v \ll c$ this ratio approaches 2, as expected. Conversely, in the ultra-relativistic limit $v \to c$ the ratio tends to 1, indicating that the net normal force vanishes. This provides a natural explanation why a photon follows a straight-line trajectory:
as already mentioned, the brachistochronic motion represents a subtle and sophisticated interplay between maximizing velocity and minimizing distance. In the classical regime, where there is no upper bound on velocity, it can be advantageous to lengthen the path in order to increase the speed, thereby reducing the overall travel time. In contrast, in the ultra-relativistic limit, where velocity is bounded from above, the optimization shifts toward minimizing the distance—i.e., effectively motion along a straight line. This result is consistent with the conclusions of the 1986 paper by Goldstein and Bende**r** [6].

**I. Stationary approach:**

Claim: A necessary and sufficient condition for a brachistochronic path is
$F_n = mv^2/R_c = 2(E-U)/R_c$ (where $R_c$ is the radius of curvature).

Proof: Brachistochronic path is stationary. The condition for a stationary motion is that up to a first order the *relative* change in length of any small section of the trajectory is equal to the *relative* change in the velocity in this sector [4]. Since if, for example, the length is shortened by 1% and the speed is reduced by 1%, the total sliding time does not change to the first order. Therefore it requires:

$$\delta S / S = \delta v / v \qquad .1$$

Here $S$ is an arbitrary *infinitesimal* segment of the brachistochronic orbit (cf. Fig 3). Let $S_1$ be the alternative tiny orbit shifted by a small amount $\varepsilon$ relative to it, as shown in Fig. 3, and $v$ is the velocity along $S$. So according to Fig. 3:

$$S = R_c \beta \quad S_1 = (R_c - \varepsilon)\beta \qquad .2$$

Here $R_c$ is the radius of curvature of that segment. The trajectory was shortened by:

$$\delta S = S - S_1 = \beta \varepsilon \quad \rightarrow \quad \delta S / S = \varepsilon / R_c \qquad .3$$

Our goal is therefore to prove that if $F_n = mv^2/R_c$, then Eq. (1) holds.

Now, the infinitesimal work required to shift the bead from $S$ to $S_1$ equals the product of the normal component $F_n = mv^2/R_c$ by the distance, $\varepsilon$, between the two tracks and it also equals the change of the kinetic energy. Using (3) we get:

$$\delta W = F_n \varepsilon = \left(\frac{mv^2}{R_c}\right)\varepsilon = \delta(\tfrac{1}{2}mv^2) = mv\delta v \quad \rightarrow \quad \frac{\varepsilon}{R_c} = \frac{\delta v}{v} \quad \rightarrow \quad \frac{\delta S}{S} = \frac{\delta v}{v} \qquad .4$$

Therefore $F_n = mv^2/R_c$ is a sufficient condition for the validity of equation (1).

On the other hand, according to Newton's law and since $\vec{f}_{net} \equiv \vec{N} + \vec{F}$,

$$f_{net,n} \equiv N - F_n = mv^2/R_c = F_n \qquad .5$$

Using Eq. 5 we can infer two rules: 1a "the rule of 2" and 1b "the mirror rule":

$$N = 2F_n \quad \rightarrow \quad \vec{N} = -2\vec{F}_n \qquad .6$$

And since: $\vec{f}_{net} \equiv \vec{N} + \vec{F} \quad \rightarrow \quad \vec{f}_{net,t} = \vec{F}_t \quad \vec{f}_{net,n} = -\vec{F}_n$

$$\left|\vec{f}_{net}\right| = \left|\vec{f}_{net,t} + \vec{f}_{net,n}\right| = \left|\vec{F}_t - \vec{F}_n\right| = \left|\vec{F}_t + \vec{F}_n\right| = \left|\vec{F}\right| \qquad .7$$

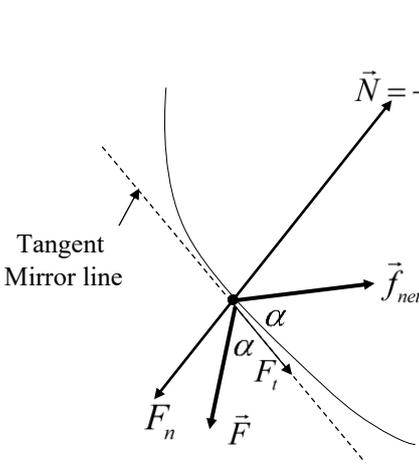

Fig. 1 In a brachistochronic track $N = 2F_n$ and $f_{net} = F$; the line of mirror is the tangent of the track.

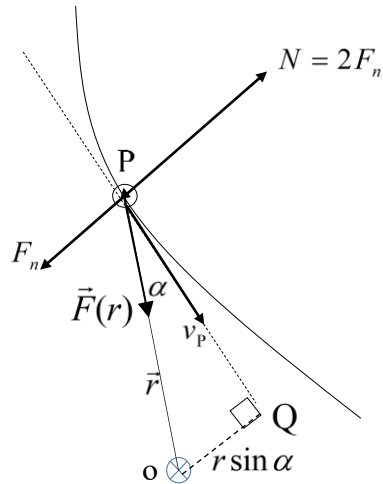

Fig. 2 The constant time $T_0$ equals the distance $r\sin\alpha$ divided by the speed $v_p$ at P.

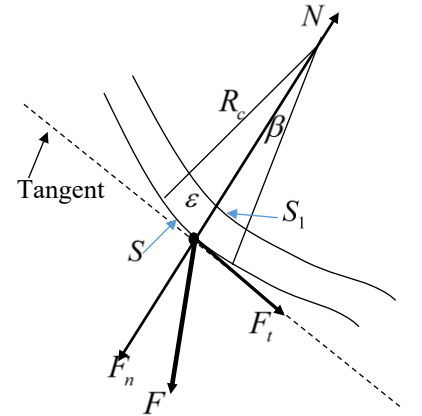

Fig. 3 In a stationary track, any tiny change does not change the total sliding time to first order.

Conclusion: throughout *any* brachistochronic motion, a net force acts on the bead whose magnitude is equal to the magnitude of the general conservative force at that point. It can be shown (cf. Fig. 1) that when the tangent to the trajectory forms an angle $\alpha$ with respect to the direction of the force, then $\vec{f}_{net}$ forms an angle $2\alpha$ with respect to the conservative force. Therefore we can say that $\vec{f}_{net}$ is a mirror image of $\vec{F}$ where the line of mirror is the tangent of the brachistochronic trajectory. We conclude that the *magnitude* of the acceleration along the path equals the *magnitude* of the conservative force divided by the mass of the bead. In appendix II, we demonstrate how the application of the mirror rule facilitates a simple and rapid solution to the Bernoulli brachistochrone problem.

### II. E-L approach

We wish to prove via the E-L equation that $F_n = mv^2/R_c$ is a sufficient condition that the E-L equation holds. Starting with geometry: from Fig. 4 below we see that the force component in the direction normal to the path exerted by a general conservative force $\vec{F}$ is

$$F_n = F_x \cos\gamma + F_y \sin\gamma \qquad .8$$

Here $\gamma$ is the angle between the tangent and the vertical. And it is given that this normal component force equals $mv^2/R_c$.

From e.g. ref [5] the E-L equation for brachistochronic trajectory is:

$$\psi y'' = (\psi_y - \psi_x y')(1+y'^2) \quad \rightarrow \quad \frac{y''}{(1+y'^2)} = \frac{(\psi_y - \psi_x y')}{\psi} \qquad .9$$

where $\psi \equiv 1/v = \left[(2/m)(E-U)\right]^{-1/2}$ and $\psi_x \equiv \partial\psi/\partial x$, etc.

Multiplying both sides in (9) by $(1+y'^2)^{-1/2} = 1/\sqrt{1+\tan^2\phi} \equiv \cos\phi = \sin\gamma$ we get:

$$\frac{y''}{(1+y'^2)^{3/2}} = \frac{(\psi_y - \psi_x y')}{\psi(1+y'^2)^{1/2}} = \frac{(\psi_y - \psi_x y')}{\psi}\sin\gamma \quad \rightarrow \quad \frac{1}{R_c} = \frac{(\psi_y - \psi_x y')}{\psi}\sin\gamma \qquad .10$$

(Since the radius of curvature is: $R_c \equiv (1+y'^2)^{3/2}/|y''|$).

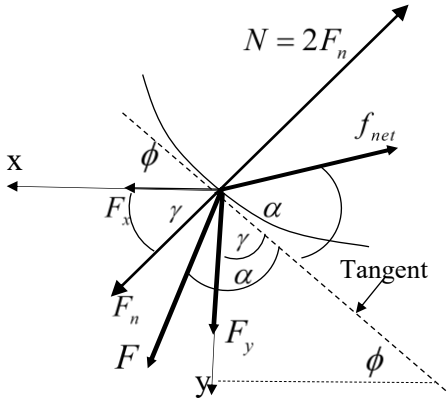

Fig. 4 The magnitude of the steering force, N, is twice as large as the magnitude of the normal component of the conservative force $F_n$.

Since $F_x = -\partial U/\partial x \quad F_y = -\partial U/\partial y$, then from the above definition of $\psi$ we get:

$$\psi_x y' = \left(\tfrac{1}{2}\sqrt{2/m}(E-U)^{-3/2}\cot\gamma\right)F_x \qquad .11$$

because the slope of the trajectory is: $y' = -\cot\gamma$.

$$\psi_y = \left(\tfrac{1}{2}\sqrt{2/m}(E-U)^{-3/2}\right)F_y \qquad .12$$

Inserting these expressions in Eq. 10 and since $\tfrac{1}{2}mv^2 = E-U$, after some algebra:

$$\frac{1}{R_c} = \frac{(\psi_y - \psi_x y')}{\psi} \sin\gamma = \frac{(F_y + F_x \cot\gamma)\sin\gamma}{2(E-U)} = \frac{F_y \sin\gamma + F_x \cos\gamma}{2(E-U)} = F_n \qquad .13$$

$$2(E-U)/R_c = mv^2/R_c = F_y \sin\gamma + F_x \cos\gamma = F_n \qquad .14$$

From this we conclude that if $F_n = mv^2/R_c$, the E-L equation holds.

The proof of rules 1a and 1b follows the same reasoning as presented in the previous section

### III. A stationary approach for a central force:

We wish to prove that in a brachistochronic motion under a *general central* force $\vec{F}(r)$ the ratio between the angular momentum and the kinetic energy is constant: $L_o / E_k = 2T_0 =$ const.
The proof is based on the fact that $dL_o / dE_k = L_o / E_k$ implies $L_o / E_k = \text{const} \equiv 2T_0$, where $T_0$ has the dimension of time. First let's calculate this ratio according to Fig. 2:

$$\frac{L_o}{E_k} = \frac{mvr\sin\alpha}{\frac{1}{2}mv^2} = 2\left(\frac{r\sin\alpha}{v}\right) \qquad .15$$

Now, from basic dynamic considerations (cf. Fig. 2) in every infinitesimal period of time $dt$ the angular momentum changes due to the torque $\tau_0$ exerted by the normal force of the wire:

$$dL_o = \tau_0 dt = rN\cos\alpha\, dt \qquad .16$$

(The central force $\vec{F}(r)$ does not exert any torque).

The kinetic energy changes due to the tangential component of the conservative force:

$$dE_k = d(\tfrac{1}{2}mv^2) = m\dot{v}v\,dt = ma_t v\,dt = m(F_t/m)v\,dt = (F\cos\alpha)v\,dt \qquad .17$$

Therefore the ratio between the above infinitesimal changes:

$$\frac{dL_o}{dE_k} = \frac{Nr\cos\alpha\,dt}{Fv\cos\alpha\,dt} = \frac{Nr}{Fv} \equiv \frac{Nr}{Fv}\cdot\left(\frac{\sin\alpha}{\sin\alpha}\right) = \left(\frac{N}{F_n}\right)\left(\frac{r\sin\alpha}{v}\right) \qquad .18$$

Since $N = 2F_n$, comparing (15) and (18) we get:

$$\frac{dL_o}{dE_k} = \frac{L_o}{E_k} = 2\left(\frac{r\sin\alpha}{v}\right) \equiv 2T_0 \qquad .19$$

Note: The constant expression $r\sin\alpha / v \equiv T_0$ has dimensions of time. One meaning of this time is deciphered by Fig. 2: $T_0$ - The time required for the bead to travel along $\overline{OQ}$ from the center of force to the continuation of the tangent when it moves at a constant speed $v_p$, the instantaneous speed at p. For an attractive force $T_0 = r_{min}/v_{max}$ and for a repulsive force $T_0 = r_{max}/v_{max}$ (since when $r$ is extremal $\alpha = 90^0$; see Figs. IV.2 and IV.4 in appendix IV and the explanation there). Further explanation of the significance of the constant $T_0$ is provided in the following section.

### IV. E-L approach for a central force:

Claim: For a central potential the quantities $T_0 = r\sin\alpha/v$ and $L_o/E_k$ are constants.

The angle $\alpha$ is defined as the angle between the tangent to the curve and the force $\vec{F}$, as shown in Fig. 1. Since the *central* force is independent of the polar angle $\theta$, the E-L equation in this case according to the Beltrami identity reduces to a first order differential equation:

$$L - r'\frac{\partial L(r,r')}{\partial r'} = \text{const.} \equiv T_0 \qquad .20$$

Here, $T_0$ is a constant that originates from the fact that the "Lagrangian" is independent of the polar angle $\theta$, and

$$L = \frac{\sqrt{r^2 + r'^2}}{v(r)} \qquad r' \equiv \frac{dr}{d\theta}$$

By substituting in equation (20), we get:

$$\frac{\sqrt{r^2 + r'^2}}{v} - \frac{1}{v}\frac{r'^2}{\sqrt{r^2 + r'^2}} = \frac{(r^2 + r'^2 - r'^2)}{v\sqrt{r^2 + r'^2}} = \frac{r^2}{v\sqrt{r^2 + r'^2}} = T_0 \qquad .21$$

Hence:

$$T_0 = \left(\frac{r}{v}\right)\left[\frac{r}{\sqrt{r^2 + r'^2}}\right] \qquad .22$$

It remains to prove that $\frac{r}{\sqrt{r^2 + r'^2}} = \sin\alpha$.

In appendix I we show that the angle $\alpha$ satisfies:

$$\tan\alpha = r/r'(\theta). \qquad .23$$

Using $\sin\alpha \equiv \tan\alpha/\sqrt{1+\tan^2\alpha}$
and substituting from equation (22), we obtain:

$$\sin\alpha \equiv \frac{\tan\alpha}{\sqrt{1+\tan^2\alpha}} = \frac{(r/r')r'}{\sqrt{r^2+r'^2}} = \frac{r}{\sqrt{r^2+r'^2}} \qquad .24$$

Finally, substituting into equation (22), we conclude that the quantity $T_0 = r\sin\alpha/v$ is indeed a constant with dimensions of time, as required.

And now, we multiply the numerator and the denominator of this expression by $mv$:

$$T_0 = \frac{r\sin\alpha}{v} = \left(\frac{mv}{mv}\right)\left(\frac{r\sin\alpha}{v}\right) = \frac{mvr\sin\alpha}{mv^2} = \frac{L_0}{2E_k} \rightarrow \frac{L_0}{E_k} = 2T_0 = \text{const.} \qquad .25$$

### V. Replacing the wire by magnetic field

The next task is to demonstrate that for a *central* force $\vec{F}(r)$, the wire can be replaced by a magnetic field, and we will explore its properties. The effect of the wire consists in its applying the normal force $\vec{N} = -2\vec{F}_n$. This may be mimicked by a magnetic field.

However, before that, we will prove that the radius of curvature of the brachistochrone path in this case satisfies the relation: $R_c = rmv/F(r)T_0$.

Using Fig. 2 and the results obtained, we have obviously:

$$\sin\alpha = vT_0/r \qquad .26$$

From Fig. 2, we derive:

$$F_n = F(r)\sin\alpha = F(r)vT_0/r \qquad .27$$

However, as we have shown, the component of the net force in the normal direction equals in magnitude and is opposite in direction to the normal component of the conservative force. Thus:

$$F_n = f_n = mv^2/R_c \qquad .28$$

From this, we derive:

$$R_c = \frac{rmv}{F(r)T_0} \qquad .29$$

We now require that the magnetic force $F_B = qvB$ replace the normal force $N$ exerted by the wire. Therefore:

$$N = qvB = 2\frac{mv^2}{R_c} \rightarrow B(r) = 2\frac{mv}{qR_c} = \frac{2mv(F(r)T_0)}{q(mvr)} = 2\left(\frac{T_0}{q}\right)\left(\frac{F(r)}{r}\right) \qquad .30$$

Assuming that the central force is generated by an electric field of magnitude $\vec{E}(r) = \vec{F}(r)/q$ we can write:

$$\vec{B}(r) = 2T_0 \left( \frac{E(r)}{r} \right) \hat{z} \qquad .31$$

Notes:
1. We assume that the brachistochronic motion occurs in a plane $(x, y)$ and that the magnetic field is oriented perpendicular to the plane of motion so $\vec{B} \perp \vec{E}$.
2. From Eq. (31), we conclude that in the case of a harmonic force $\vec{F}(r) = \pm K\vec{r}$ the magnetic field has a constant magnitude. In all other cases, generally the magnetic field will depend on the distance $r$.
3. It is well known that in the case of a constant force (a limiting case of a central force when the center of force is at an infinite distance), the magnetic field is constant, too.
4. As required the divergence of this magnetic field is zero $\vec{\nabla} \cdot \vec{B} = 0$.
5. The constant $2T_0$ has the dimensions of time and may serve as a calibration factor determined by the initial conditions.
6. It is surprising to discover that *any* charged particle placed in a general *central* electric field and a corresponding magnetic field, as defined in Eq. (31), will follow a brachistochronic trajectory for the force $\vec{F}(r) = q\vec{E}(r)$, similar to a massive body moving in a gravitational field along a geodesic path of minimal length.
7. Obviously even if $F(r)$ is not of electromagnetic origin, but $B$ is defined through eq. 30 (with $q = 1$ to have the correct units)), the resulting curve will be a brachistochrone for $F(r)$.

## VI. Advantages of Using a Magnetic Field Instead of a Wire
When calculating the brachistochronic trajectory, the objective is to determine the shape of the smooth wire that guides the particle to its destination. This involves solving the E-L nonlinear differential equation. However, substituting the wire by a magnetic field, as described in Eq. (31), simplifies the problem significantly. Bypassing the E-L equation, the task is shifted to the solution of equations of motion of a charged particle in electromagnetic field which is straightforward and computationally efficient as shown in appendix IV. To this end, we developed a short, dedicated MATLAB code to address the problem. The core of the software consists of only seven lines of code (appendix IV). During the process of calculating the trajectory, we also compute the acceleration and velocity at every point along the path. As a result, upon completing the calculation, we acquire additional tools that allow us to confirm easily that the computed trajectory is indeed brachistochronic, satisfying:

a. The "rule of 2": $\vec{N} = -2\vec{F}_n$

b. The "mirror rule": $|\vec{F}| = |\vec{f}| \rightarrow |\vec{F}_t + \vec{F}_n| = |f_t + f_n| = |\vec{F}_t - \vec{F}_n| = m|\vec{a}|$

## Summary

In this paper, we have shown that a given trajectory is indeed a brachistochrone under a specified conservative force if and only if the magnitude of the normal component of this force equals the corresponding centrifugal force. Two equivalent rules follow from this result:
1. The rule of 2
2. The mirror rule
Two additional equivalent rules are valid only for a central force:
3. The ratio between angular momentum and kinetic energy is constant.
4. The time). Fig. 2is a constant quantity ( $T_0 = r\sin\alpha / v$

A consequence of these rules reveals that in the case of a central force, the wire can be effectively replaced by a magnetic field whose form can be explicitly determined. This substitution allows one to bypass the use of the E-L equation and instead solve the equations of motion directly, in a fast and accurate manner.

Moreover appendix IV demonstrates that during the computation of the brachistochrone path based on the equations of motion, additional information is obtained regarding the particle's velocity and acceleration. This information is significant, as it supports the assertion that the path is indeed a brachistochrone.

## **Acknowledgment**

We thank Prof. Amos Ori for his very significant contributions to the research through original ideas, advice, and encouragement.

## **Appendix I:** Finding the angle between the tangent and the radius vector

We calculate the angle $\alpha$ between the tangent and the radius vector $\vec{r}$ of the curve $r(\theta)$. From Fig I.1:

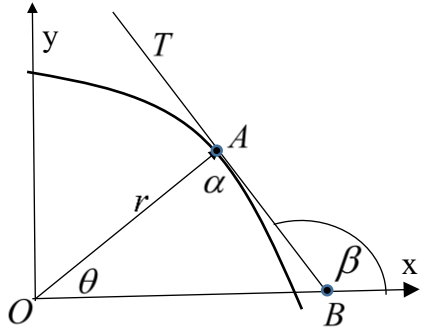

Fig. I.1 $\beta = \alpha + \theta$ is an external angle of the triangle AOB

$$x = r\cos\theta \;\rightarrow\; dx = (r'\cos\theta - r\sin\theta)d\theta \quad (r' \equiv dr/d\theta)$$
$$y = r\sin\theta \;\rightarrow\; dy = (r'\sin\theta + r\cos\theta)d\theta$$
I.1

At point A, the slope of the tangent is given by:
$$\tan\beta = \frac{dy}{dx} = \frac{r'\sin\theta + r\cos\theta}{r'\cos\theta - r\sin\theta} \qquad \text{I.2}$$

Dividing both numerator and denominator by $r'\cos\theta$ we obtain:
$$\tan\beta = \frac{\tan\theta + (r/r')}{1 - (r/r')\tan\theta} \qquad \text{I.3}$$

But $\beta$ is external angle of the triangle AOB hence
$$\beta = \theta + \alpha \;\rightarrow\; \tan\beta = \tan(\theta + \alpha) = \frac{\tan\theta + \tan\alpha}{1 - \tan\theta\tan\alpha} \qquad \text{I.4}$$

Equating this expression with (I.3) we derive:

$$\tan\alpha = r/r' \quad\rightarrow\quad \alpha = \tan^{-1}(r/r') \qquad \text{I.5}$$

$$\sin\alpha \equiv \frac{\tan\alpha}{\sqrt{1+\tan^2\alpha}} = \frac{r/r'}{\sqrt{1+(r/r')^2}} = \frac{r}{\sqrt{r^2+r'^2}} \qquad \text{I.6}$$

.

**Appendix II: A Simple Solution of the Classical Brachistochrone Problem**

The purpose of this appendix is to demonstrate that, by relying on mirror the rule it is straightforward to prove that the brachistochrone path under a constant force *mg* is a cycloid. Also, we will see how the smooth wire can be replaced by a constant magnetic field.

<u>Proof</u>: The total energy is conserved: $\tfrac{1}{2}m\dot{x}^2 + \tfrac{1}{2}m\dot{y}^2 = mgy_0$ and from the mirror rule it follows that the magnitude of the acceleration at every point along the path is *g*.

Thus $\sqrt{\ddot{x}^2 + \ddot{y}^2} = g$. Since the motion is periodic* (like a pendulum), we assume a solution in the form of harmonic functions**.

$$\ddot{y}(0) = g \quad \ddot{x}(0) = 0 \quad\rightarrow\quad \ddot{y} = g\cos\omega t \quad \ddot{x} = g\sin\omega t \qquad \text{II.1}$$

Here $\omega$ is some constant. For simplicity, we place the starting point at the origin (0, 0) and the endpoint at (L, 0). Initial conditions $x(0) = y(0) = 0 \quad \dot{x}(0) = \dot{y}(0) = 0$.

From this, by performing two time integrations, we obtain the cycloid curve:

$$x = a(\omega t - \sin\omega t) \quad y = a(1 - \cos\omega t) \qquad \text{II.2}$$

It is straightforward to show that this solution satisfies the law of energy conservation, subject to the aforementioned initial conditions. The particle reaches its destination when $\omega t = 2\pi$ and then $x(2\pi) = L = 2\pi a \quad\rightarrow\quad a = L/2\pi$ and $a \equiv g/\omega^2 \quad\rightarrow\quad \omega = \sqrt{g/a}$.

Now, if we want to get identical path by replacing the smooth wire with a constant magnetic field, we will require, according to the "rule of 2", that the magnetic force be twice the normal force exerted by gravity. For simplicity, we choose the lowest point of the path (at $y = 2a$) and charge the particle with a charge *q*. And we will require that the following condition holds: $F_B = qv_{\max}B_0 = q\sqrt{4ga}\,B_0 = 2mg \quad\rightarrow\quad B_0 = \left(\dfrac{m}{q}\right)\sqrt{\dfrac{g}{a}} = \omega\left(\dfrac{m}{q}\right)$

---

\* When dealing with a central force, it is, in principle, always possible to extend a brachistochrone path originating from a point A located at a distance *a* from the force center, such that the path reaches a point B situated at the same distance. Consequently, a bead released from rest at point A will reach point B, momentarily come to a stop, and then return to its point of origin. In other words, it will execute a periodic motion akin to that of a pendulum.

\*\* At $t = 0$, the particle's velocity is zero, and thus it must move vertically downward with an acceleration *g*. If the motion were to start with an angle other than zero with respect to the vertical, its net acceleration would have differed from *g*, which would contradict the mirror rule.

# Appendix III Brachistochronic Path in a Radial Electric Field $\vec{E} = -K\vec{r}$ :

In this appendix, we will prove, using the "mirror rule": $|\vec{F}| = |\vec{f}| = m|\vec{a}|$, that the brachistochronic path of an *attractive* harmonic force $\vec{F} = -K\vec{r}$ is a hypocycloid.

Initial conditions: $x(0) = a \quad y(0) = 0 \quad \dot{x}(0) = \dot{y}(0) = 0$

It must hold that:

$$|\vec{a}| = \sqrt{\ddot{x}^2 + \ddot{y}^2} = (K/m)r = \omega_0^2\sqrt{x^2 + y^2} \rightarrow \ddot{x}^2 + \ddot{y}^2 = \omega_0^4(x^2 + y^2) \quad \omega_0^2 \equiv K/m \qquad \text{III.1}$$

Since the motion is *periodic*, we assume a solution in the form of harmonic functions, subject to the initial conditions:

2.III $\qquad\qquad\qquad x(t) = A\cos(\omega_1 t) + B\cos(\omega_2 t) \qquad y(t) = A\sin(\omega_1 t) - B\sin(\omega_2 t)$

It follows:

$A + B = a \quad \omega_1 A - \omega_2 B = 0 \rightarrow B = (\omega_1/\omega_2)A \qquad$ Hence:

$$A = a\omega_2/(\omega_2 + \omega_1) \quad B = a\omega_1/(\omega_2 + \omega_1) \qquad \text{III.3}$$

We choose $\omega_1 = 1(\text{rad./s})$ and define: $b \equiv a \cdot 1/(\omega_2 + \omega_1) = B$ and get: $A = a - b$

Therefore, substituting into the proposed solution in (III.2) yields the well-known hypocycloid equations in the Cartesian coordinate system:

$$x(t) = (a-b)\cos(t) + b\cos(\tfrac{a-b}{b}t) \qquad y(t) = (a-b)\sin(t) - b\sin(\tfrac{a-b}{b}t) \qquad \text{III.4}$$

Note: Similarly, it can be proven that the shape of the brachistochronic path in a central *repulsive* harmonic force $\vec{F} = +K\vec{r}$ is an epicycloid

$$x(t) = (a+b)\cos(t) - b\cos(\tfrac{a+b}{b}t) \quad y(t) = (a+b)\sin(t) - b\sin(\tfrac{a+b}{b}t).$$

# Appendix IV Notes on the Calculation of the Trajectory Using MATLAB

This appendix presents the core of the MATLAB code developed for trajectory computation. The code consists of only seven lines, structured as follows:

1. The first two lines compute the acceleration components: $a_x, a_y$
2. The next two lines compute the velocity components $v_x, v_y$.
3. The subsequent two lines calculate the trajectory component $x, y$.
4. The final line computes the resultant velocity $v = \sqrt{v_x^2 + v_y^2}$.

The code is written for a central field of the form $E(r) = K/r^n$. However, with minor modifications, it can handle any central electric field, such as:
$E(r) = \pm Kr^n, E(r) = \pm E_0 \exp(\pm kr), \ E(r) = E_0 \cosh(kr)$ or similar cases.

```
Ax(k) = (1/r(k-1)^(n+1))*(-wc*x(k-1) - Vy(k-1)*wb);
Ay(k) = (1/r(k-1)^(n+1))*(-wc*y(k-1) + Vx(k-1)*wb);
Vx(k) = Vx(k-1) + Ax(k)*dt;
Vy(k) = Vy(k-1) + Ay(k)*dt;
x(k)  = x(k-1) + Vx(k)*dt;
y(k)  = y(k-1) + Vy(k)*dt;
V(k)  = sqrt(Vx(k)^2 + Vy(k)^2);
```

Here: $wb \equiv \omega_b = qB/m \quad wc \equiv \omega_c = Cq/m$ (C is a constant)

The software output includes graphs of all relevant quantities. Moreover, based on the criteria developed in this paper, it provides graphical verification that the presented trajectory is indeed a brachistochronic path as shown in Figs. IV.1-IV.4.

Fig. IV.1 demonstrates that the required condition is satisfied, confirming $F_B \equiv qvB = 2F_n$ this trajectory as such. Fig. IV.2 presents a brachistochronic trajectory under an attractive Coulomb force $F(r) = -K/r^2$. Fig. IV.3 and IV.4 illustrate the corresponding case for a repulsive force of the form $F = +Kr^3$.

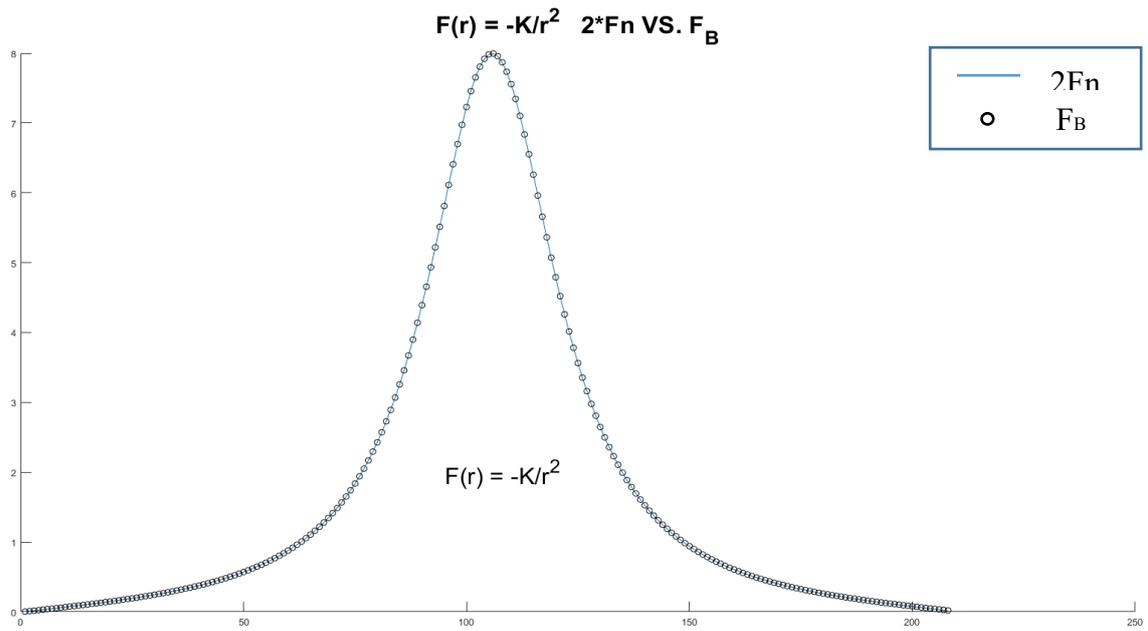

Fig. IV.1 From the "rule of 2", it follows that the magnitude of the magnetic force is twice the magnitude of the normal component of the conservative force. $F_B = 2F_n \rightarrow \vec{F}_B = -2\vec{F}_n$

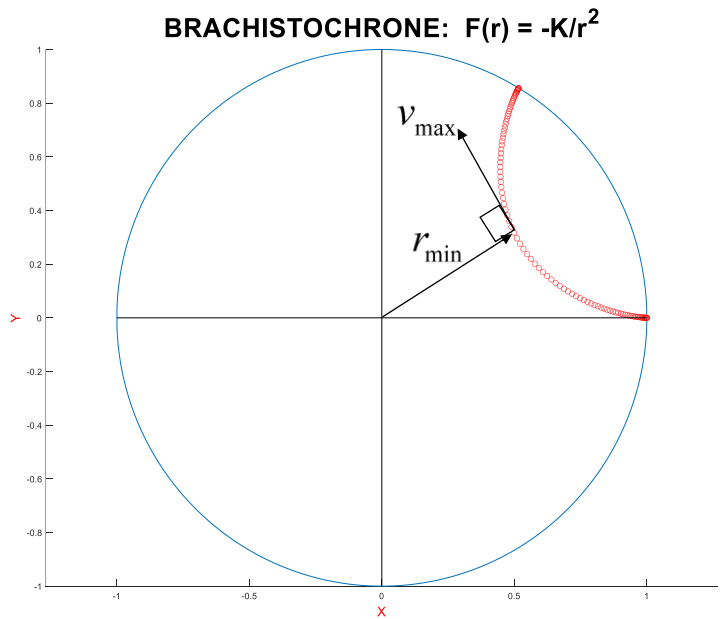

Fig. IV.2 Brachistochrone trajectory (in red) under an attractive Coulomb force. $\vec{F}(r) = -K/r^2$
Note that $r_{min}$ is perpendicular to the trajectory and to the velocity. Hence: $T_0 = r_{min}/v_{max}$

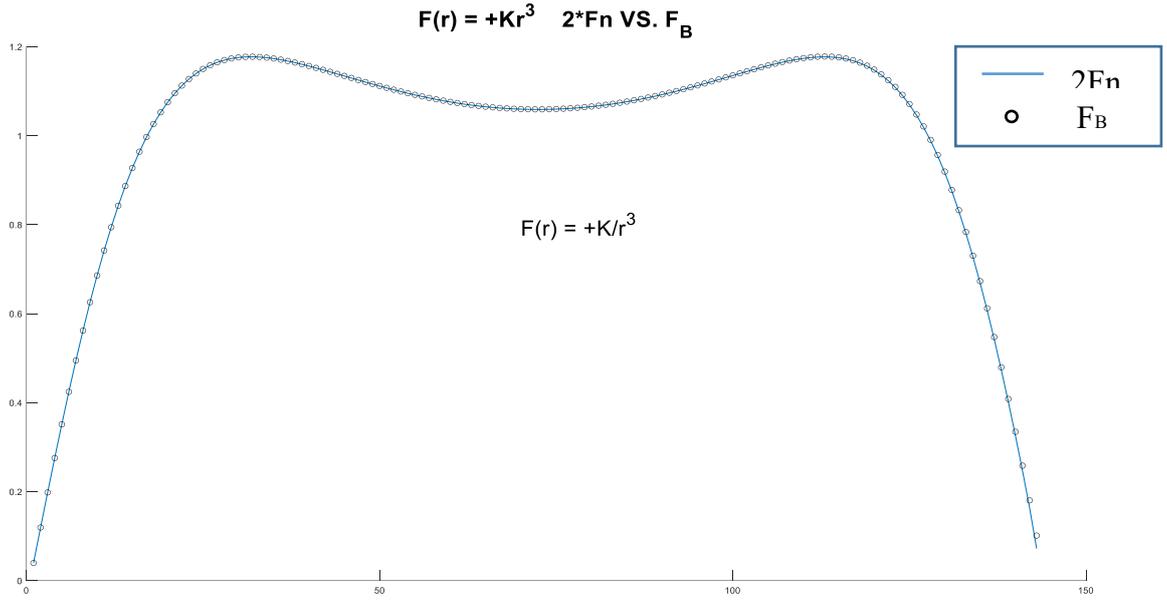

Fig. IV.3 From the "rule of 2", it follows that the magnitude of the magnetic force is twice the magnitude of the normal component of the conservative force. $F_B = 2F_n \rightarrow \vec{F}_B = -2\vec{F}_n$

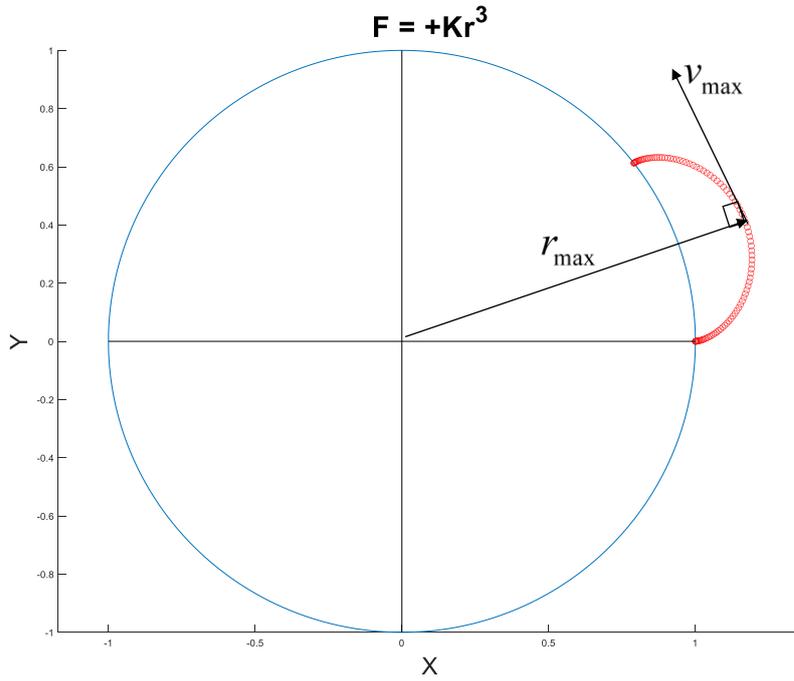

Fig. IV.4 Brachistochrone trajectory (in red) under a repulsive force. $\vec{F}(r) = +Kr^3 \hat{r}$. Note that $r_{max}$ is perpendicular to the velocity and to the path. Hence: $T_0 = r_{max} / v_{max}$